# Electronic transport descriptors for the rapid screening of thermoelectric materials

Tianqi Deng[1]*, Jose Recatala-Gomez[2,3]*, Masato Ohnishi[4]*, D. V. Maheshwar Repaka[3]*, Pawan Kumar[3], Ady Suwardi[3], Anas Abutaha[3,5], Iris Nandhakumar[2], Kanishka Biswas[6], Michael B. Sullivan[1], Gang Wu[1], Junichiro Shiomi[4#], Shuo-Wang Yang[1#], Kedar Hippalgaonkar[3,7#]

[1] Institute of High Performance Computing, Agency for Science, Technology and Research (A*STAR), 1 Fusionopolis Way, Singapore 138632, Republic of Singapore

[2] Department of Chemistry, University of Southampton, University Road, Highfield, Southampton SO17 1BJ, United Kingdom

[3] Institute of Materials Research and Engineering, Agency for Science, Technology and Research (A*STAR), 2 Fusionopolis Way, Singapore 138634, Republic of Singapore

[4] Department of Mechanical Engineering, The University of Tokyo, Hongo 7-3-1, Bunkyo-ku, Tokyo 113-8656, Japan

[5] Qatar Environment and Energy Research Institute, Hamad Bin Khalifa University, Qatar Foundation, Doha, 34110, Qatar

[6] New Chemistry Unit and School of Advanced Materials and International Centre for Materials Science, Jawaharlal Nehru Centre for Advanced Scientific Research (JNCASR), Bangalore 560064, India

[7] School of Material Science and Engineering, Block N4.1, 50 Nanyang Avenue, Nanyang Technological University, Singapore 639798, Republic of Singapore

#Correspondence to: kedar@ntu.edu.sg, yangsw@ihpc.a-star.edu.sg, shiomi@photon.t.u-tokyo.ac.jp

*: These authors contributed equally to this work

**Abstract**

The discovery of novel materials for thermoelectric energy conversion has potential to be accelerated by data-driven screening combined with high-throughput calculations. One way to increase the efficacy of successfully choosing a candidate material is through its evaluation using transport descriptors. Using a data-driven screening, we selected 12 potential candidates in the trigonal $ABX_2$ family, followed by charge transport property simulations from first principles. The results suggest that carrier scattering processes in these materials are dominated by ionised impurities and polar optical phonons, contrary to the oft-assumed acoustic-phonon-dominated scattering. Combined with calculations of thermal conductivity based on three-phonon scattering, we predict *p*-type $AgBiS_2$ and $TlBiTe_2$ as potential high-performance thermoelectrics in the intermediate temperature range for low grade waste heat harvesting, with a predicted *zT* above 1 at 500 K. Using these data, we further derive ground-state transport descriptors for the carrier mobility and the thermoelectric power factor. In addition to low carrier mass, high dielectric constant was found to be an important factor towards high carrier mobility. A quadratic correlation between dielectric constant and transport performance was established and further validated with literature. Looking ahead, dielectric constant can potentially be exploited as an independent tuning knob for improving the thermoelectric performance.

**Introduction**

The advent of machine learning (ML) and high-throughput (HT) density functional theory (DFT) computation has shifted the scientific process from a time consuming Edisonian approach to a more efficient, *in-silico* approach.[1–3] The deployment of these tools has led to multiple advancements: prediction of novel compounds, either by HT-DFT[4] or ML,[5] that were later on realized experimentally resulting in a knock-on effect of an acceleration of materials diagnosis,[6] and the fast screening of promising material candidates[7–10] using materials descriptors, *i.e.* features that are inherent to the material, easily calculated and have a direct relationship with a functional property.[11]

One widely used method for *ab initio* calculation of charge transport properties is the Boltzmann transport equation (BTE).[12] The commonly adopted constant scattering time approximation (CSTA), *i.e.*, energy independent scattering time, is well-known for its simplicity. In this method, a single, constant value of scattering time ($\tau_0$) is assigned to all charge carriers.[13] However, such an inherent assumption is not always accurate in reality and depends on the arbitrary choice of $\tau_0$.[14,15] A commonly adopted approximation that bypasses CSTA's shortcomings is based on the deformation potential theory (DPT)[16], that adequately describes the long-wavelength intra-band electron-acoustic-phonon scattering for non-polar semiconducting materials.[15,17–19] More recently, an electron-phonon averaged (EPA) approximation has been introduced as an alternative approximation, which includes both acoustic and optical phonon scattering and was applied to half-Heusler alloys.[14] Nonetheless, the long-range Fröhlich-type scattering by polar optical phonon (POP), which is particularly important for polar compound semiconductors,[20] is not included in DPT or EPA approximations. A recently developed approach, Energy-dependent Phonon- and Impurity-limited Carrier Scattering Time AppRoximation (EPIC STAR), achieves good accuracy for polar materials at lower computational cost and is therefore appropriate for HT screening of such materials.[21] Currently, the Wannier interpolation technique[22,23] allows for accurate electron-phonon calculations and has become the state-of-the-art method for the prediction of phonon-limited, charge transport properties.[4,24–27] Yet, scattering time computation remains computationally expensive due to large number of Brillouin zone sampling points needed for numerical integration.[24,25] This is especially critical for polar materials, where numerical integration becomes even more difficult.[28–30] Therefore, an easy-to-compute descriptor could facilitate rapid initial screening without performing complex computation for materials with lower predicted potential.

A study of HT-DFT computations combined with first principles transport calculations was reported in 2008 by Yang *et. al.*[13] Their screening selected 36 potential half-Heusler candidates and the thermoelectric properties of these compounds were evaluated by means of BTE in CSTA. They proposed LaPdBi as a new *n*-type half-Heusler material with potential thermoelectric applications.[13] Also in the half-Heusler family and based on the work reported by Gautier *et. al.*[31], Zhou *et. al.*[4] conducted accurate EPW calculations on 15 compounds, which led them to explain the large power factor in NbFeSb and ZrNiSn. They also concluded that optical phonons are the dominant scattering mechanism for charges in many half-Heuslers, in good agreement with available experimental data.[4] Zhu *et. al.*[32] performed stability studies on 27 compounds belonging to the $V^1$-VIII-$V^2$ family (with $V^1$ = V, Nb, and Ta; VIII = Fe, Ru, and Os; and $V^2$ = As, Sb, and Bi) of half-Heuslers, predicting six compounds to be stable. This was further verified via synthesis and optimization the TaFeSb-based half-Heuslers, reporting a peak *zT* of ~1.52 at 973 K for the $Ta_{0.74}V_{0.1}Ti_{0.16}FeSb$ alloy.[32]

Outside the half-Heusler family, Li *et. al.* screened the Materials Project (MP) database in order to study $ABX_2$ compounds whose thermoelectric properties were unexplored.[33] They conducted HT-DFT calculations on 41 candidates whose band gap fell in the range of 0.1-2.5 eV and predicted that 12 of them, both *n*- and *p*-type, have high figure of merit (*zT*). Further, some were experimentally realized, like $CuInTe_2$, demonstrating the potential of these approaches.[33] A similar study was conducted by Xi and co-workers.[15] They screened the Materials Informatics Platform (MIP) and, selected 214 inorganic compounds out of 82412 by only looking at select cation/anion combinations in a FCC anion lattice with cation coordination number = 4. Further, looking at those that have bandgap values >0.1 eV, electrical transport calculations were performed using DPT, assuming that the electron−phonon coupling is insensitive to band variations. One of the candidates, $Cd_2Cu_3In_3Te_8$, was experimentally realized and *zT*~1 was found at approximately 900 K.[15] From the above studies, it is clear that incorporating relevant scattering rates in charge transport calculations is necessary to estimate the electrical transport properties accurately. However, there is a long-standing lack of general transport property descriptors for scattering mechanisms beyond DPT, especially as acoustic phonon scattering may not be the dominant scattering mechanism in many thermoelectric materials at optimal doping.

Herein, we leverage upon the richness of the Materials Project (MP) Database[34], screening for high symmetry, low band gap[35] chalcogenide compounds. From the MP Database, we focus on the trigonal (space group number 166, $R\bar{3}m$) $ABX_2$ family where A are monovalent elements from the alkali and transition metals (Na, K, Sc, Cr, Ag), lanthanoids (Gd) and group V (Tl), B are trivalent elements from the lanthanoids (Gd), transition metals (Cr, Au), groups III and V (In, Bi, Sb, Tl) and X are chalcogenides (S, Se, Te). $ABX_2$ compounds were chosen due to their tendency to have low thermal conductivity, with the possibility of $ns^2$ lone pair electrons,[36] which enabled us to narrow down our focus on charge transport properties. Our findings reveal that in this family, polar optical phonon scattering, which has been neglected often in literature, is significant even with heavy doping. Our calculations further single out the p-type compounds $AgBiS_2$ and $TlBiTe_2$ with potentially high TE performance and *zT* above 1 at 500 K in both cases. More importantly, we proposed charge transport descriptors based on ground state properties and easy-to-obtain parameters. The descriptors described herein qualify themselves as robust first level thermoelectric screening parameters, which obviate the need for computationally expensive calculations. We expect this strategy to be widely implemented in the quest for high-performance inorganic TE materials in a wide temperature range.

## *Materials informatics and candidate screening*

We establish a screening strategy for the identification of unexplored potential TE candidates. To do so, we first make use of appropriate material descriptors for the rapid assessment of key properties directly correlated to the performance.[37] First, we screen for compounds with bandgaps between 0.16 and 4 eV, as it relates to the maximum Seebeck coefficient that can be achieved at a particular temperature ($S_{max} \sim \frac{E_g}{2eT_{max}}$).[38] To screen for stable compounds that will not decompose into different crystal structures at a fixed composition, we set a strict threshold for the energy above the convex Hull of zero; the thermodynamic stability as indicated by the energy above convex Hull is evaluated at 0 K. At finite temperature (T > 0 K), however, the contribution from configurational entropy has been reported to stabilize compounds with $E_{Hull} > 0$. In other words, a small, non-zero $E_{Hull}$ at 0 K does not necessarily render experimental synthesis impossible ($E_{Hull} < 80$ meV).[39,40] This is confirmed by looking at the most recent entry for trigonal $AgBiS_2$ in the MP database (mp-29678), which now shows $E_{Hull} \sim$ 20 meV and has been experimentally realized.[41] Next, the number of charge carrying valleys, or the valley degeneracy ($N_v$) is key to achieve high Seebeck coefficient and electrical conductivity simultaneously, and is preferentially found in high symmetry structures.[42] We set a threshold for compounds with more than four symmetry operations, thus increasing the probability of having compounds with high $N_v$ present in the dataset. Finally, to ensure data sufficiency, we screened for binary, ternary and quaternary chalcogenides. Domain knowledge motivates the choice of chalcogenide materials: traditionally, chalcogenides are good TE materials.[43–49] Our screening resulted in nearly 600 compounds (combining binary, ternary and quaternary), from which we focused on ternary chalcogenides, as they represent the majority fraction in the mined data. In the end, the initial dataset was comprised by 146 ternary chalcogenides with chemical formula $ABX_2$ (where $A^{1+}$ and $B^{3+}$ are cations and $X^{2-}$ is the chalcogen anion) and trigonal structure (space group $R\bar{3}m$). This initial dataset was reduced to 12 compounds after filtering out the low performing candidates based on previously calculated CSTA power factors from Ricci *et al.*,[50] to compare CSTA with our detailed scattering time calculations. In addition, it is noteworthy to mention that TE properties have only been experimentally reported for 3 out of 12 compounds from our dataset (trigonal $TlBiTe_2$, trigonal $TlSbTe_2$ and cubic $AgBiS_2$), leaving an unexplored chemical space.[51–55] *Ab initio* transport property simulations were then performed for these 12 compounds, and a pictorial representation of the computational framework deployed in this work can be found in Figure 1.

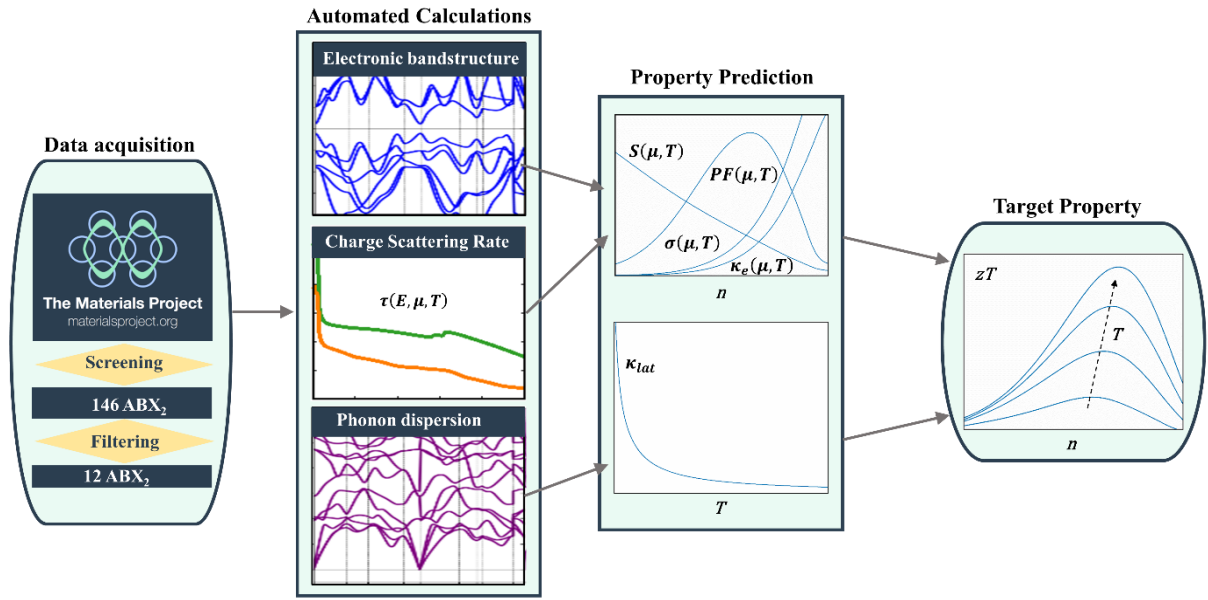

**Figure 1.** Computational framework deployed in this work. The electron band structure and DFPT phonon calculations of the 12 candidate compounds were carried out using QUANTUM ESPRESSO. Afterwards, charge transport properties were simulated using the EPIC STAR method, taking into consideration scattering events produced by both acoustic and optical phonons, ionized impurities and Thomas Fermi screening of free carriers.[21] Lattice thermal conductivities were calculated using phonon BTE implemented in ALAMODE, including 3-phonon scattering events.[56]

## Results and Discussion

Figure 2(a) shows the crystal structure for $ABX_2$ compounds, with space group $R\bar{3}m$ (No. 166). The layered structure is comprised by [B-X] slabs in the *b-c* plane with A cations orthogonal to this plane (along the *c*-axis). The atoms A and B occupy octahedral positions and interact with X with dissimilar strength, depending on the ion charge and specific position they occupy in the slab. On the other hand, the X atoms are octahedrally coordinated with respect to AB. To illustrate the bonding nature and crystal symmetry, Figure 1(b) shows the DFT band structure for $AgBiS_2$, which has an indirect bandgap of ~ 0.65 eV, where the valence band maxima is between the Γ and X high symmetry points and the conduction band minima is centred at the Z point. According to the partial density of states (PDOS), the valence band maxima is mainly comprised of sulphur p-states with modest contribution from silver d-states, whilst the conduction band minima is mainly bismuth p-states, with minor contributions from silver and sulphur. The phonon dispersion shown in Figure 2(c) is comprised of three acoustic branches and nine optical branches, with the lowest optical phonon located at *ca.* 50 cm$^{-1}$. The proximity of the optical branches to the acoustic branches is expected to lead to a decreased lattice thermal conductivity near room temperature, due to increased scattering phase space and higher likelihood of phonon-phonon scattering. The phonon density of states in Figure 2(c) shows the optical branches at low energy mainly comprised of vibrations of Ag atoms, followed by vibrations of Bi atoms. Additional verification is given by the atomic participation ratio (APR), which quantifies the degree of participation of different atoms in a specific phonon mode.[57] We observe that Ag and Bi have a large participation ratio (red colour) in the phonon modes of the lowest optical branches. In addition, these low-lying optical branches are flat and avoid crossing the acoustic branches at certain high symmetry points of the Brillouin zone (for instance at the Γ and L points). The combination of high participation ratio, flat low energy optical branches and avoided crossing with acoustic branches hints at localized phonon vibration which would potentially decrease the lattice thermal conductivity.[58] Further, $ABX_2$ compounds have attracted interest due to the presence of lone-pair electrons, which are expected to result in softened phonon modes. Note that experimentally synthesized $AgBiS_2$ results in the rock-salt disordered cubic Fm3-m space group, different from our study of the R3-m space group. In rock-salt compounds with the $ABX_2$ formula, the presence of $ns^2$-orbitals[59] induces structural instabilities that translate into an increase of anharmonicity in the bonding, ultimately resulting in strong phonon-phonon interactions that can potentially reduce the lattice thermal conductivity as low as the amorphous limit.[36] Figure 2(d) shows a bar plot representing the theoretical thermoelectric performance for a range of temperatures, from 300 K to 900 K, for seven out of the twelve $n$- and $p$-type ABX$_2$ chalcogenides, (note that thermal conductivity calculations are more computationally expensive, hence were not performed on all 12 compounds). The thermoelectric figure-of-merit $zT_{max}$ was calculated using lattice thermal conductivity in the amorphous limit, while $zT_{cryst}$ was calculated using lattice thermal conductivity corresponding to perfectly crystalline samples (Figure S1(a)). In addition, we also calculated $zT$ using the lattice thermal conductivity for polycrystalline samples with grain size of 1 μm (Figure S1(a)) and the resulting $zT$ border the values of the single crystal samples ($zT_{cryst}$). This is because the phonon mean free path is much smaller than 1 μm (see Figure S1(b)), so the impact of grain boundaries becomes less relevant. In many cases, $p$-type ABX$_2$ compounds show higher thermoelectric performance than their $n$-type counterpart, due to a high band degeneracy which is attributed to their complex valence band structure.[9]

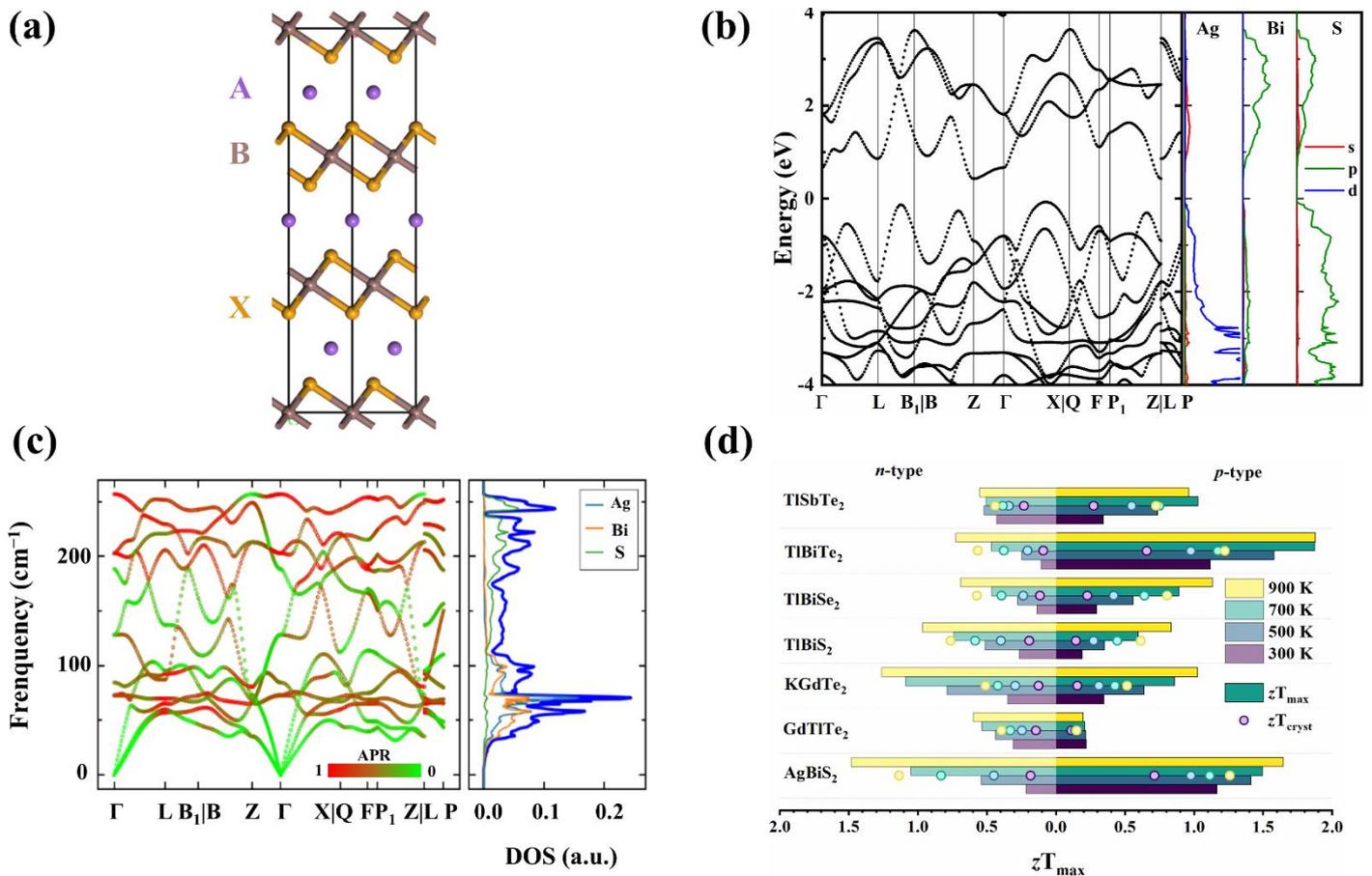

**Figure 2.** (a) Atomic structure of ABX$_2$ chalcogenides, (b) Calculated electronic band structure for AgBiS$_2$, (c) Calculated phonon dispersion for AgBiS$_2$. The colour of the bands denotes atomic participation ratio (APR), ranging from green (zero, no participation in the phonon mode) to red (one, large participation in the phonon mode). (d) Maximum $zT$ ($zT_{max}$, bar graph) and $zT$ for the single crystal sample ($zT_{cryst}$, circles) as a function of temperature for selected $n$- and $p$-type ABX$_2$ chalcogenides. $zT_{max}$ has been calculated employing the minimum lattice thermal conductivity (amorphous limit) while $zT_{cryst}$ has been calculated for a single crystal by including three-phonon scattering.

Among these, the $p$-type AgBiS$_2$ and TlBiTe$_2$ compounds could reach a $zT$ above 1 at room temperature. Especially, the predicted value for $n$-type AgBiS$_2$ is higher than the experimental results obtained by Rathore *et al.*[41] This mismatch is likely because the optimal carrier concentration ($1.48 \times 10^{19}$ cm$^{-3}$ for $n$-type AgBiS$_2$ at room temperature) was not experimentally realized, and their synthesis resulted in the cation-distorted cubic rock salt structure.[41] Also, their analysis concluded that further optimization of the carrier concentration through doping in $n$-type AgBiS$_2$ was required to achieve better performance. Interestingly, our calculations show the $p$-type TlBiTe$_2$ compound is expected to achieve a maximum $zT$ of ~1.9 at 900K, which is much larger than previously reported experimental values for the material (0.15 at 760K), signifying the potential for further optimization.[51] However, within the ABX$_2$ family, the best performance is attained by the cubic $p$-type alloy AgSbTe$_{1.85}$Se$_{0.15}$, with $zT \sim 2$ in the temperature range 550-600 K, mainly due to further reduction in the thermal conductivity from point defects and stacking faults.[60] In fact, this material is also cubic (space group $F\bar{3}dm$) as opposed to the trigonal (space group $R\bar{3}m$) ABX$_2$ compounds studied here, and it is not currently contained in the Materials Project database, which explains its absence from the potential candidate dataset resultant from the screening.[60]

We also study in detail the representative charge scattering mechanism that makes these compounds good prospects for mid-temperature thermoelectric applications. Figure 3 shows the energy dependence of the scattering rate for both $n$- and $p$-type AgBiS$_2$ and TlBiTe$_2$ at 300 K for the optimal doping condition (determined from the peak of the powerfactor, S$^2\sigma$ as a function of carrier concentration).

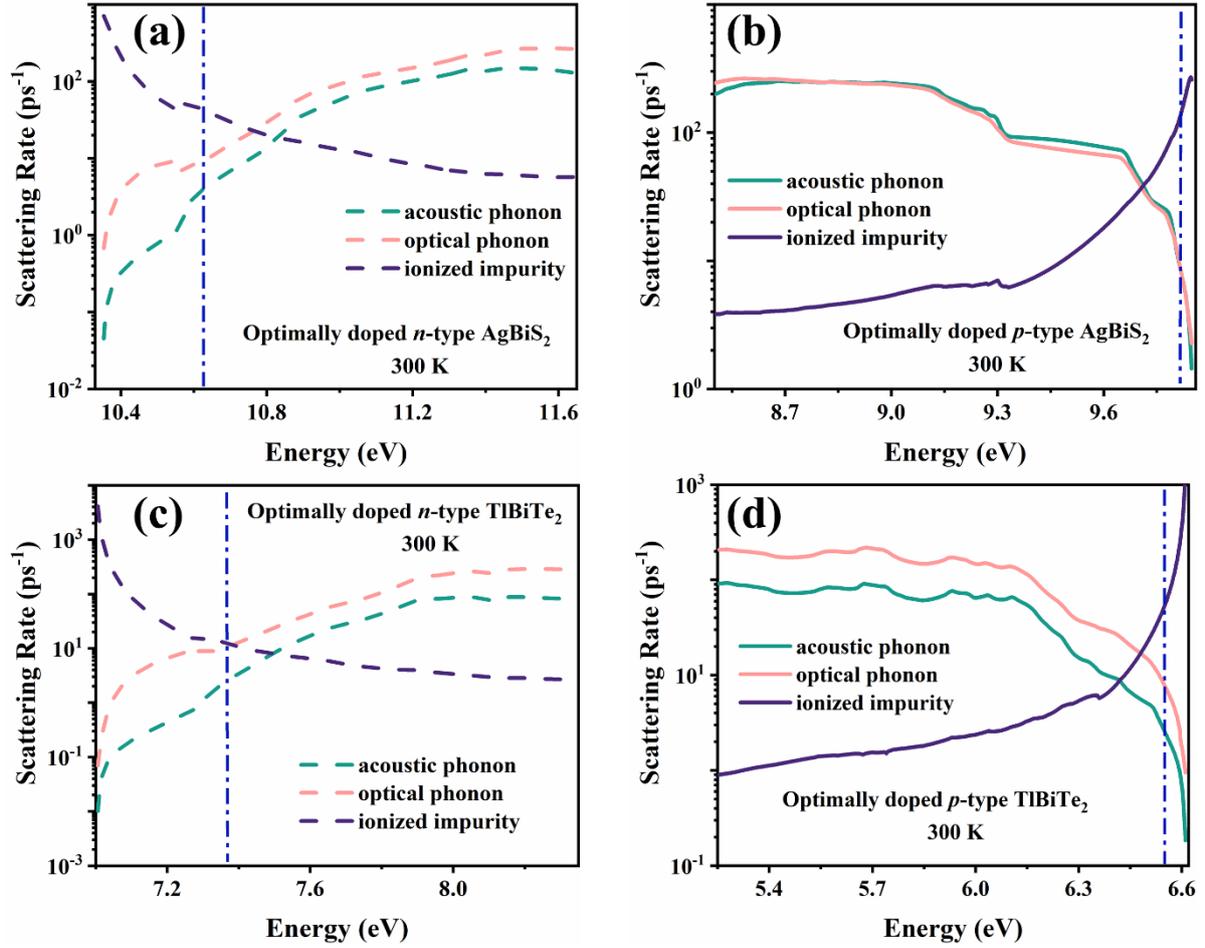

**Figure 3.** Room temperature scattering rate for optimally doped (a) *n*-type and (b) *p*-type AgBiS$_2$ and (c) *n*-type and (d) *p*-type TlBiTe$_2$. The vertical blue line indicates the Fermi level at optimal doping level.

In general, the same scattering phenomena are dominant, from the energy dependencies, for both *n*-type and *p*-type materials at 300 K. For these optimally doped materials, ionised impurities dominate the charge scattering around the Fermi level (blue dash-dotted lines in Figure 3). Interestingly, even in such a heavily doped regime where charge carrier scattering arising from polar optical phonons is partially screened by the free carriers, we still observed a strong contribution from optical phonons, that surpasses the acoustic phonon contribution. The contribution from optical phonons is especially significant for *n*-type TlBiTe$_2$ [Figure 3(c)], comparable to that from ionised impurities. Moreover, for high-energy carriers the optical phonon scattering even dominates over ionized impurities. Similar trends are observed for the *p*-type materials, as shown in Figure 3(b) and 3(d). This emphasizes the importance of polar optical phonon scattering, even in the heavily doped case where free-carrier screening is strong. In fact, in both *n*-type and *p*-type AgBiS$_2$ and TlBiTe$_2$, the overall scattering rate has a higher contribution from the optical phonons as compared to the acoustic phonons, which is in stark contrast to the widely used assumption of acoustic-phonon-dominated scattering in the literatures.[61,62] This signifies the importance of including the polar optical phonon (POP) scattering contribution for polar materials, as this could potentially indicate the dominance of POP despite high temperature and doping. Consequently, using acoustic phonon limited assumption in analyzing the charge transport properties can result in substantial error, particularly in ABX$_2$ as well as half Heusler class of compounds.[4]

In light of this understanding, and to facilitate the rapid screening and identification of potential high-performance thermoelectric materials, key descriptors which account for these scattering mechanisms can be deduced. Previously the quality factor *B* proposed by Wang *et al.* has been adopted in the screening for high performance thermoelectrics.[63] Nevertheless, this requires one key assumption: the scattering is dominated by acoustic phonons throughout the doping and temperature range under study. However, in our case of ABX$_2$ compounds, and possibly in other potential thermoelectric materials, the *B*-factor would not qualify as a reasonable descriptor, as from Figure 3 we observed that the charge scattering events are controlled by polar optical phonons and ionized impurities.

To derive such descriptors, we first observe that the POP scattering rate is proportional to the Fröhlich coupling strength $\tau_{POP}^{-1}(E) \propto |C_{POP}|^2$, which in turn can be written as a function of Born effective charge, phonon displacement and dielectric constant. Importantly, it is inversely proportional to the dielectric constant via $C_{POP} \propto \frac{1}{\hat{q}\cdot\varepsilon\cdot\hat{q}}$.[20,64] Therefore, the POP scattering time should also be proportional to dielectric constant squared $\tau_{POP}(E) \propto \varepsilon^2$. Similarly, the ionized impurity scattering strength also depends on the dielectric constant as the charges also experience electrostatic screening.[21,65] Therefore in the Brooks-Herring model, the scattering rate from ionized impurity is also inversely proportional to dielectric constant squared $\tau_{IIS}(E) \propto \varepsilon^2$. Since the dielectric constant determines the overall electrostatic interaction, we conjecture that the interaction strength for other phonons may also be inversely correlated to $\varepsilon$. Therefore, the overall scattering rate should be strongly correlated to $\varepsilon^2$ (further details in Supplementary Section 1). Thus, we propose a general transport descriptor by considering mixed scattering contribution from ionized impurities and polar optical phonons. The descriptor for the carrier mobility is obtained by applying the relation $\mu = \frac{e\tau}{m_C^*}$ considering that the total scattering time is proportional to $\varepsilon^2$:

$$\mu \propto \varepsilon^2 \, m_C^{-1} \qquad (1)$$

where $\mu_{avg}$ is the direction averaged mobility at optimal carrier concentration, $m_c$ is the conductivity effective mass, $\varepsilon$ is the dielectric constant and $T$ is the absolute temperature in K.

Figure 4(a) shows the correlation between the transport descriptor and the direction averaged mobility for our calculated $ABX_2$ compounds. We benchmarked our data together with experimental values from the literature, in order to validate the descriptor.[43–47,49,66] The temperature dependent effective masses, conductivity effective masses and the dielectric constant were used when reported[44,46,47], whereas for other compounds we used values reported in the Landolt–Börnstein database, e.g. for effective mass for PbTe[67], or other literature e.g. for the dielectric constant of PbTe.[68] . Interestingly, materials of different crystal systems (*e.g.* cubic PbTe and orthorhombic SnSe) follow the same trend, hinting that the descriptor could be generally applicable.

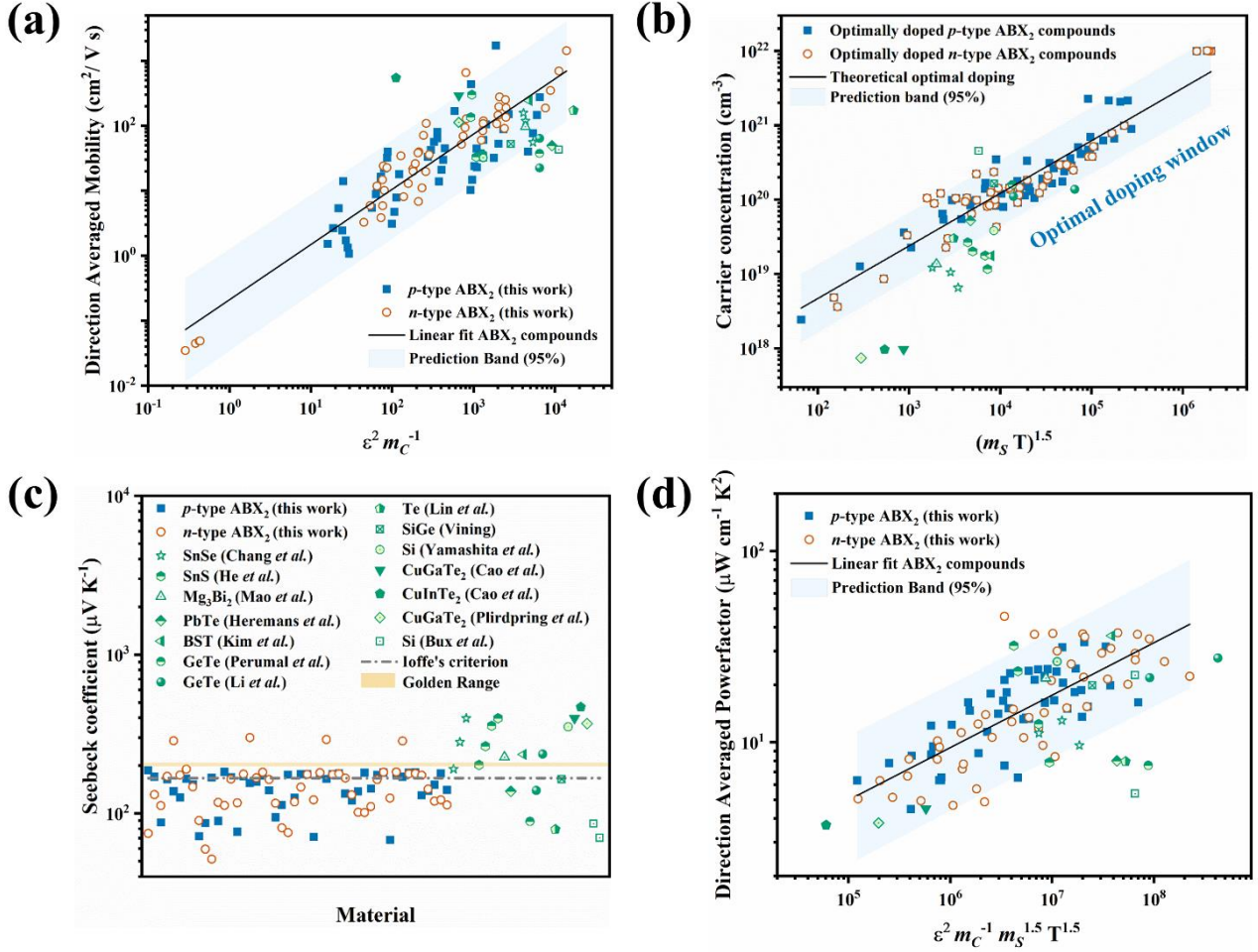

**Figure 4.** Transport descriptors for ABX$_2$ materials. (a) Direction averaged mobility transport descriptor at optimal doping. (b) Optimal carrier concentration (n$_{opt}$) descriptor.[69] (c) Seebeck coefficient for each material (x-axis, ABX$_2$ materials are listed in alphabetical order) versus theoretical criteria (Ioffe's criterion[70] and golden range[71]). (d) Direction averaged power factor transport descriptor at optimal doping. The benchmark corresponds to experimental data.[43–47,49,66,72–74,75,76,77] The data was taken from the respective references and averaged using the method described by Parker et al.[78] The blue area in (a), (b) and (d) corresponds to the prediction band calculated for ABX$_2$ compounds. It corresponds to the range of values that are likely to contain the value of a new observation, with a 95% confidence.

In additional to the carrier mobility, another key quantity in optimizing the thermoelectric performance is the carrier concentration. Assuming a single carrier type, the optimal carrier concentration (n$_{opt}$) should be proportional to the Seebeck effective mass and the temperature, $n_{opt} \propto (m_S T)^{1.5}$.[69] In Figure 4(b), this trend is plotted for *n*- and *p*-type ABX$_2$ compounds as well as for the literature values used for benchmark, at their corresponding carrier concentration. We observe that, as expected, our compounds follow the trend whilst some literature values deviate from this ideal relationship, indicating that the reported carrier concentration may not be at the optimal level.

Next, a descriptor for the power factor (PF = S$^2\sigma$; PF $\propto$ S$^2 n\mu$) is derived, where the mobility descriptor is given by Equation 1 and a descriptor for optimal carrier concentration is given by $n_{opt} \propto (m_S T)^{1.5}$.[69] However, a descriptor for the Seebeck coefficient is not as readily available. The underlying difficulty of finding a descriptor for the Seebeck coefficient resides in the fact that theoretically, the Seebeck coefficient at optimal doping should be a constant.[79] This originates from the steady-state solution to the BTE, that states that the optimal Seebeck coefficient only depends on reduced Fermi potential and scattering exponent and thus, is independent of effective masses, the valley degeneracy and scattering strength.[65] Hence, under the parabolic band approximation, the power factor will be maximized at a single value of Seebeck coefficient, S$_{opt}$. This was first proposed by Ioffe, who reported that the optimized value for Seebeck coefficient is 172 µV K$^{-1}$.[70] Later, Pichanusakorn and Bandaru showed that S$_{opt}$ can be found in a range of 130-187 µV K$^{-1}$ and that the most frequent value was 167 µV K$^{-1}$.[80] Recently, Hong *et al.* expanded on this issue by reporting that the optimized Seebeck coefficient is not a single value but a range that changes depending on whether we are optimizing *zT* or the power factor. The authors reported that to achieve maximum power factor, the Seebeck values range from 195 µV K$^{-1}$ to 202 µV K$^{-1}$.[71] The Seebeck tendency to accumulate around a range of values has also been noted by Zhang *et al.*[81] Figure 4(c) shows the Seebeck coefficient of our ABX$_2$ compounds along with the Seebeck coefficient of the literatures. Deviations from the optimal value may be indicative of non-optimal doping. However, we observe that while the majority of theoretical and experimental values of S approach the theoretically predicted limits, they still span a wider range.

Nevertheless, we introduce a transport descriptor for the direction averaged power factor at optimal doping (PF$_{avg}$), shown in Equation 3. They key assumption is that the carrier concentration is given by its optimal value (n = n$_{opt}$ $\propto$ (m$_S$ T)$^{1.5}$), while the Seebeck coefficient is around an optimal constant (S ~ S$_{opt}$ ~ constant):

$$PF \propto \varepsilon^2 \, m_C^{-1} \, m_S^{1.5} \, T^{1.5} \qquad (3)$$

The comparison between $PF_{avg}$ and the descriptor are shown in Figure 4(d). We indeed observe an increasing trend of $PF_{avg}$ with respect to the transport descriptor for both simulated *n*- and *p*-type ABX$_2$ compounds as well as for the literature, with some literature values deviating from the general trend. This is a consequence of assuming optimal doping during the derivation of the PF descriptor: the power factor of materials with non-optimal doping will be lower than the maximal power factor that can be achieved. This also provides a theoretical guidance for experimental optimization of thermoelectric performance via tuning the carrier concentration towards optimal doping level. The good correlation between the transport descriptors and the transport properties potentially enable a facile method for first-level screening of potential TE candidates from easy to calculate parameters. In order to compare to the CSTA derived descriptors in Ricci *et al.*, we also observed that another descriptor for the powerfactor, the Fermi surface complexity factor ($N_V^*K^*$, Figure S2)[9] indeed also captures the trend. Importantly, our descriptor is related to $N_V^*K^*$, as it captures the band features, but contains more information regarding the scattering rate, which is absent in $N_V^*K^*$ due to the CSTA assumption.

In addition, we also analyze phonon properties of these ABX$_2$ compounds to gain insights into their phonon anharmonicity and instability, which lead to low thermal conductivities (Figure S3(a)). In order to explore anharmonicity, Nielsen *et. al.*[36] used an applied electric field and displacement of atoms, that significantly deform the lone-pair charge density of the group V element, resulting in the structural instability and strong phonon anharmonicity of ABX$_2$ compounds. For materials composed of guest atoms and a framework such as skutterudites[82,83] and clathrates[84], effects of phonon anharmonicity can be analyzed by comparing phonon properties of the pristine structure and structure excluding the guest atoms. One, however, cannot employ this approach for ABX$_2$ compounds because removing the group V elements breaks the structure. In this analysis, we have applied hydrostatic strains to ABX$_2$ compounds to tune its phonon anharmonicity.

Because the three-phonon scattering is a complicated process, we have carefully analyzed change of harmonic and anharmonic terms on phonon relaxation times when such a strain was applied to the system. Here, we analyzed AgBiS$_2$, which exhibits promising electronic power factor. As shown in Figures 2(c) and 5(a), ABX$_2$ compounds have flat bands at low frequency. The most intuitive effect of the flat band may be enhancement of Scatting Phase Space (SPS). SPSs of absorption (+) and emission (-) processes, $P_3^\pm(q) = \left(\frac{1}{N}\right)\sum_{q_1 q_2} \delta(\omega \pm \omega_1 - \omega_2)\, \delta(\boldsymbol{q} \pm \boldsymbol{q}_1 - \boldsymbol{q}_2 - \boldsymbol{G})$, have been computed with and without including the effect of the lowest optical branch, one of the flat bands around 60-80 cm$^{-1}$ shown as a bold line in Figure 5(a). This flat band increases SPS of absorption (emission) process at frequencies lower (higher) than its frequency as shown in Figure 5(b). It is worth noting that while the lone-pair electrons result in whole features of atomic vibrations rather than only in flat bands, the flattening is one of the representative features of the lone-pairs resulting in weak bonding.

While the presence of flat bands may be related to the phonon anharmonicity and local distortions, SPS is a harmonic phonon property. We have, therefore, applied a hydrostatic strain, a uniform expansion, to the compound to explore its phonon anharmonicity. As shown in Figure 5(a), transverse acoustic (TA) modes around T point, $\boldsymbol{q}$ = (0.5, 0.5, 0.5), are significantly modified by the applied strain; their frequencies decrease and finally approaches imaginary values, which are represented as negative values in Figure 5(a). The instability of TA modes on Γ-X line can be clearly confirmed by the Grüneisen parameter, defined as the change in the frequency with the crystal volume, as shown in Figure 5(c). Red markers in Figure 5(c) show values for TA modes on $\boldsymbol{q}$ = (q, q, q), where q is an arbitrary number, and the inset clearly shows that the value increases steeply as the T point is approached. Furthermore, the instability becomes stronger at the accelerating rate under the strain as shown in the right panel of Figure 5(c). It is also intriguing to see the eigenvector of the TA mode at T point. In the TA mode at T point, bismuth atoms whose 6s$^2$ orbitals form lone-pair electrons largely move along the in-plane direction while silver atoms do not move and sulfur atoms move only slightly (see Figure S3), which may result in strong distortion of charge distribution.

Because the changes in phonon dispersion and Grüneisen parameter clearly show that the TA mode at T point is unstable and should have a strong anharmonicity, we analyze its three-phonon interaction in detail. The detailed observation of the $|V_3|^2$ term described in the *Methods* section clearly shows that this unstable phonon mode mainly interacts with phonon modes from the flat bands around 60-80 cm$^{-1}$ (see left panel in Figure S3(a)), and this interaction is enhanced by the applied strain as shown in Figure 5(d). Although SPS of the TA mode is also changed because of the change in its frequency with applied strain, its change is superseded by the change in the $|V_3|^2$ term as shown in Figure S3. Enhancement of $|V_3|^2$ terms of low frequency modes due to flat bands has been also observed in clathrate compounds. These analyses reveal complicated but intriguing effects of the flat band; it enhances not only SPS but also the three-phonon matrix elements, $|V_3|^2$, and furthermore is related to the phonon instability at low frequency. Overall, the class of compounds has many intriguing possibilities with the low-lying optical modes providing enhanced phonon anharmonic scattering, with polar optical phonon scattering being dominant for majority carriers, therefore resulting in higher electronic power factor. Thus, the expected $zT$ for optimally doped $ABX_2$ compounds is > 1 at intermediate to high temperatures.

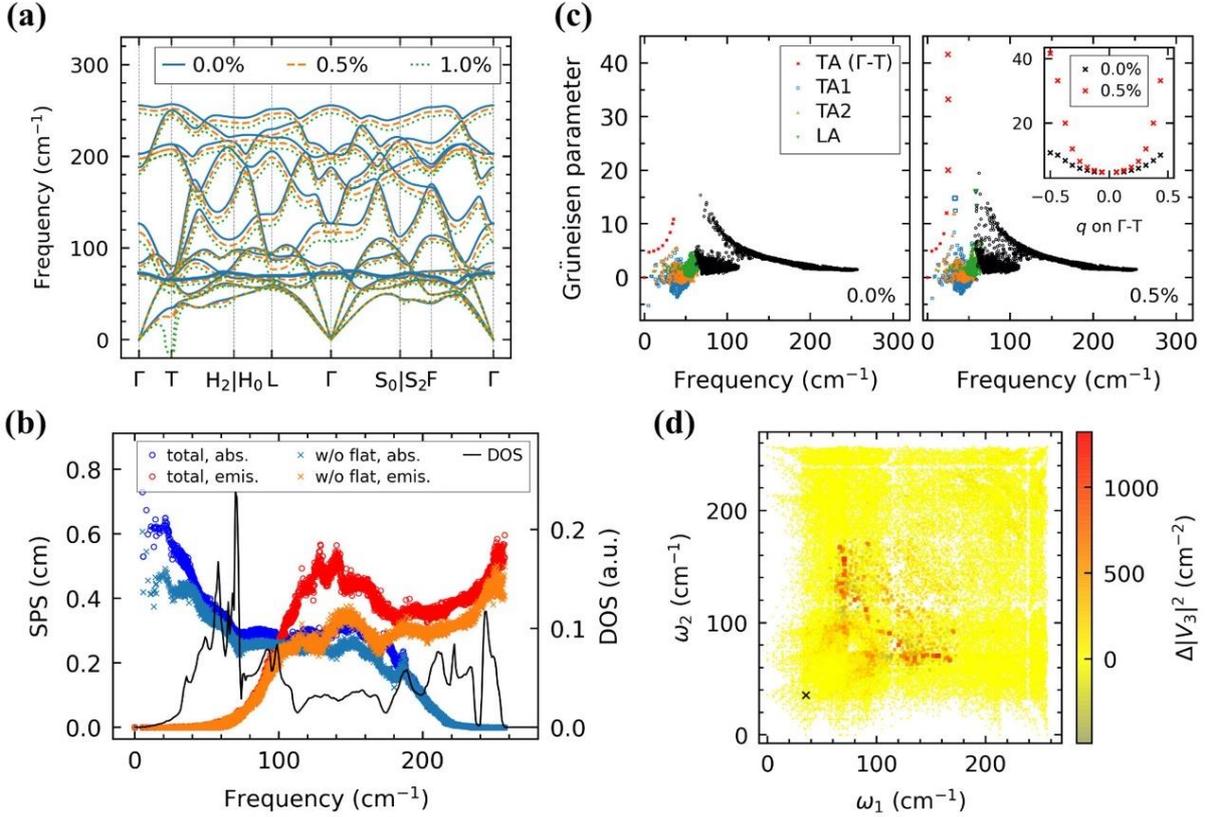

**Figure 5.** Phonon instability of $AgBiS_2$. (a) Phonon dispersions under hydrostatic strains (0.0, 0.5, and 1.0%). The transverse acoustic (TA) mode at T point ($q$ = (0.5, 0.5, 0.5)) becomes unstable under the applied strain. The bold line around 60-80 cm$^{-1}$ shows the lowest optical mode that significantly enhances phonon scattering. (b) Contribution of the flat band to scattering phase space (SPS). Blue and red markers show data for absorption and emission processes while circles and crosses show, respectively, the total value and the value excluding the effect of the lowest optical mode. Black line shows phonon density of states (DOS). (c) Grüneisen parameters of the pristine structure (left) and the structure under 0.5% strain (right). Red crosses show data for the TA modes along $q = (q, q, q)$, where $q$ is an arbitrary number, corresponding to the Γ-T line and their maximum value corresponds to the T point. Inset shows data for the TA mode on Γ-T line with respect to $q$. (d) Change in the $|V_3|^2$ term with the 0.5% strain for the TA mode at T point, which is marked with a cross (x) in the bottom panel.

## Conclusions

Two novel transport descriptors for the rapid screening of potential thermoelectric materials with high mobility and power factor have been introduced. Pre-existing information in Materials Project database was used to filter out 12 potential candidates, and their charge transport properties have been calculated and used to derive the descriptors. We report two $p$-type compounds (AgBiS$_2$, and TlBiTe$_2$) that could achieve $zT$ larger than unity above room temperature, and a maximum figure-of-merit of 1.88 for TlBiTe$_2$ at 900 K, higher than previously reported values. Inspection of the charge carrier scattering rates for this family of compounds reveals that in heavily doped regime, while ionised impurities have a dominant scattering contribution, polar optical phonon scattering is also important and non-negligible, and must be considered for screening of new thermoelectric materials. Transport descriptors for carrier mobility and power factor are proposed by including this new insight and validated with the literature. In addition to effective mass, we proposed that dielectric constant plays an important role in determining the carrier mobility and the maximum power factor. Excellent agreement with theoretical and experimental data is observed, hence validating its use as first-level screening parameter in the search of novel materials. In addition, the anharmonic scattering terms have been explicitly considered to study the phononic thermal conductivity and AgBiS$_2$ and TlBiTe$_2$ emerge as promising candidates for intermediate temperature thermoelectrics.

**Methods**

The *ab initio* charge transport calculations were carried out using an Energy-dependent Phonon- and Impurity-limited Carrier Scattering Time AppRoximation (EPIC STAR), a fast and reliable first principles method based on density functional perturbation theory (DFPT) phonon calculation.[21] QUANTUM ESPRESSO was used for DFPT calculation and the electron density of states and group velocities were computed using BoltzTraP.[85] The charge transport properties were simulated by considering the following effects in carrier scattering: the electron-phonon scattering from both acoustic and optical, polar and nonpolar phonons, the electron-impurity scattering by ionized impurities, and Thomas Fermi screening by free carriers.[21] Generalized gradient approximation with exchange-correlation functional introduced by Perdew, Burke and Ernzerhof (PBE)[86] was used. Pseudopotentials provided by Standard solid-state pseudopotentials (SSSP)[87] with their recommended cut-off energies. 6×6×6 **k**- and **q**-point samplings are used for all compounds in their primitive cell calculations.

Lattice thermal conductivities were computed using phonon BTE implemented in ALAMODE[56], where three-phonon scattering mechanisms were included in the phonon relaxation time calculation.

Considering the second-order perturbation within the single-mode relaxation approximation, the linewidth due to the three-phonon scattering for phonon mode $q$, where $q = (\boldsymbol{q}, \omega)$ with $\boldsymbol{q}$ being the wavevector and $\omega$ being the phonon frequency is derived as

$$\Gamma(\omega) = \frac{\pi}{16} \sum_{q_1,q_2} |V_3(-q, q_1, q_2)|^2 \times [(n_1 + n_2 + 1)\delta(\omega - \omega_1 - \omega_2) - 2(n_1 - n_2)\delta(\omega - \omega_1 + \omega_2)], \quad (4)$$

where the subscripts ($i$ = 1, 2) denote phonon modes contributing to the scattering of the target mode $q$, $n_i = 1/\{\exp(\beta\hbar\omega_i) - 1\}$ is the Bose-Einstein distribution, $\beta = k_B T$ with the Boltzmann constant $k_B$, $\hbar$ is the reduced Planck constant, $N$ is the number of $\boldsymbol{q}$ points, and $-q = (-\boldsymbol{q}, \omega)$. The three-phonon matrix element $V_3$ is given by

$$V_3(q, q_1, q_2) = \left(\frac{\hbar}{N\omega\omega_1\omega_2}\right)^{1/2} \times \sum_{R_i l_i p_i} \psi^{p_0 p_1 p_2}_{0 l_0, R_1 l_1, R_2 l_2} \times \frac{e^{p_0}_{l_0}(q) e^{p_1}_{l_1}(q_1) e^{p_2}_{l_2}(q_2)}{\sqrt{M_{l_0} M_{l_1} M_{l_2}}} \times \exp[i(\boldsymbol{q} \cdot \boldsymbol{R}_0 + \boldsymbol{q_1} \cdot \boldsymbol{R}_1 + \boldsymbol{q_2} \cdot \boldsymbol{R}_2)], \quad (5)$$

where $\boldsymbol{R}_i$ is the position of the primitive cell, $l_i$ is the atom site, $p_i$ is the direction of the displacement of atom $l_i$, $M$ is the atomic mass, $\Psi$ is the cubic IFCs, and $e(q)$ is the eigenvector of the mode $q$. The phonon relaxation time $\tau_{\text{ph}}$ is given by $\tau_{\text{ph}}(q) = 1/(2\Gamma(\omega))$.

The structural optimization and computation of interatomic force constants (IFCs) were conducted with first-principles calculations using Vienna Ab initio Simulation Package (VASP).[88] PBEsol exchange-correlation functional[89], which reproduced the lattice constant well, was employed for the phonon transport analysis. Because of strong structural instability of materials[90], the structural optimization of the primitive cell needed to be carefully performed with $40 \times 40 \times 40$ $\boldsymbol{k}$-points including $\Gamma$ point; (a) the structural optimization was performed for structures with slightly different volumes, (b) the crystal volume was determined by a parabolic fitting with respect to the volume and minimized energy, and (c) the structural optimization was again performed for the structure with the optimal volume. The error of the finally-obtained minimum energy and the maximum force on the atoms were confirmed to be less than 0.01 meV and 0.01 meV/Å, respectively. IFCs were computed with a $4 \times 4 \times 1$ supercell of rectangular conventional cell which contains 192 atoms with a finite-displacement method. We have confirmed that the phonon frequency took a positive value in the whole reciprocal space. Using the obtained IFCs, we have calculated phonon properties such as scattering phase space (SPS), relaxation time, and lattice thermal conductivities ($\kappa_{\text{lat}}$). Thermal conductivity was calculated with $16 \times 16 \times 16$ $\boldsymbol{q}$-points and phonon scattering due to natural isotopes was considered with Tamura model.[91] Note that we discuss phonon transport properties of the $xx$ component in this analysis for simplicity though ABX$_2$ compounds have slightly anisotropic lattice thermal conductivity. For example, $\kappa_{\text{lat}}^{zz}$ was 17% smaller than $\kappa_{\text{lat}}^{xx}$ for AgBiS$_2$ at 300 K, where the superscript denotes the component of $\kappa_{\text{lat}}$.